# Ferroelectricity in epitaxially strained rhombohedral $ZrO_2$ thin films

J. P. B. Silva[a], R. F. Negrea[b], M. C. Istrate[b], S. Dutta[c], H. Aramberri[c], J. Íñiguez[c,d], F. G. Figueiras[e], C. Ghica[b], K. C. Sekhar[f], A. L. Kholkin[g]

[a]Centre of Physics of University of Minho and Porto (CF-UM-UP), Campus de Gualtar, 4710-057 Braga, Portugal
[b]National Institute of Materials Physics, 105 bis Atomistilor, 077125 Magurele, Romania
[c]Materials Research and Technology Department, Luxembourg Institute of Science and Technology (LIST), 5 avenue des Hauts-Fourneaux, L-4362, Esch/Alzette, Luxemburg
[d]Department of Physics and Materials Science, University of Luxembourg, Rue du Brill 41, L-4422 Belvaux, Luxembourg
[e]IFIMUP & Department of Physics and Astronomy, Sciences Faculty, University of Porto, Rua do Campo Alegre, 687, 4169-007 Porto, Portugal
[f]Department of Physics, School of Basic and Applied Science, Central University of Tamil Nadu, Thiruvarur-610 101, India
[g]Department of Physics, CICECO-Aveiro Institute of Materials, University of Aveiro, 3810-193 Aveiro, Portugal

*Electronic e-mail: josesilva@fisica.uminho.pt

## Abstract

Zirconia and hafnia based thin films have attracted tremendous attention in the last decade due to their unexpected ferroelectric behavior at the nanoscale, which facilitates the downscaling of ferroelectric devices. The present work reports a novel ferroelectric rhombohedral phase of $ZrO_2$ that can be achieved in thin films grown on (111)-Nb:$SrTiO_3$ substrates by ion-beam sputtering. Structural and ferroelectric characterizations reveal that the (111)-oriented $ZrO_2$ films are under epitaxial compressive strain and display a switchable ferroelectric polarization of about 20.2 $\mu C/cm^2$ with a coercive field of 1.5 MV/cm. Moreover, the time dependent polarization reversal characteristics of Nb:STO/$ZrO_2$/Au film capacitors exhibit bell-shape curves, a typical feature of ferroelectric films associated with domains reversal. The estimated activation field is comparable to the coercive field obtained from polarization-electric field hysteresis loops. Interestingly, the studied films show ferroelectric behavior *per se*, i.e., there is no need to apply the wake-up cycle that is essential to induce ferroelectricity in the conventional (orthorhombic) ferroelectric phase of $ZrO_2$. Therefore, the present films have a technologically advantage over the previously studied ferroelectric $ZrO_2$ films, and may be attractive for nanoscale ferroelectric devices.

**1. Introduction**

In the field of ferroelectric materials, downsizing has been a research topic for more than 20 years, until in 2011, when nano-scale ferroelectricity in doped $HfO_2$ was demonstrated [1]. Recently, the thickness range for the feasible ferroelectric performance was decreased to ~1 nm [2]. In addition to undoped $HfO_2$, $ZrO_2$, $Hf_xZr_{1-x}O_2$ (HZO) and $HfO_2$ doped with Si, Zr, Al or Gd have been shown to be ferroelectric as well [3,4]. These studies established that a high-pressure orthorhombic phase with non-centrosymmetric space group $Pbc2_1$, induced in a constrained environment, is the origin of ferroelectricity; and various factors – such as surface energy, stress and dopant level – have been proposed to be responsible for ferroelectricity in $HfO_2$- and HZO-based thin films [5,6].

In effect, these materials attracted great attention because of many advantages, such as large polarization density in ultrathin films, large band gap energy (>5 eV), and compatibility with CMOS (complementary metal-oxide-semiconductor) processing. Apart from the application in ferroelectric random-access memories (FeRAMs), the HZO unconventional ferroelectric materials are also promising candidates for future high-density, low-power energy-related applications based on pyroelectric energy harvesting, electrocaloric cooling, and electrostatic energy storage. However, an initial wake-up pre-cycling is needed to induce ferroelectricity in orthorhombic HZO films, which is technologically inconvenient [7].

In 2018, Wei *et al*. reported an unprecedented rhombohedral (R3m or R3) ferroelectric phase in epitaxially-strained $Hf_{0.5}Zr_{0.5}O_2$ thin films deposited on (001)-oriented $La_{0.7}Sr_{0.3}MnO_3/SrTiO_3$ substrates, suggesting that the ferroelectric rhombohedral phase is probably stabilized by a combination of the internal pressure inside the nanoparticles (due to a large surface energy) and the substrate-imposed compressive strain [8]. In this phase, the ferroelectric behavior was achieved *per se* without any wake-up cycling. Further, the effect of substrates like GaN (0001) and Si (111) on the phase formation in $Hf_{0.5}Zr_{0.5}O_2$ films has been investigated [9-11]. These studies reported a pure rhombohedral phase in 6 nm $Hf_{0.5}Zr_{0.5}O_2$ epitaxial layers grown on hexagonal GaN buffered Si substrates, whereas mixed monoclinic and rhombohedral phases formed in $Hf_{0.5}Zr_{0.5}O_2$ epitaxial layers grown directly on Si (111).

Following these studies, Zhang and co-workers employed density-functional-theory (DFT) calculations to investigate structural and polarization properties of the R3m $HfO_2$ phase; they found that reduction of the film thickness and in-plane compressive strain play a key role in stabilizing the R3m phase leading to a robust ferroelectricity of (111)-oriented R3m $HfO_2$ [12].

In this work, we focus on pure $ZrO_2$ thin films, and show that it is possible to prepare a polar rhombohedral (R3m) ferroelectric phase in epitaxially-strained films deposited on (111)-oriented Nb:$SrTiO_3$ substrates. We demonstrate that $ZrO_2$ is crystallized at a lower temperature than $HfO_2$, which could be highly beneficial to the CMOS process integration [13]. Atomic-resolution TEM observations and DFT calculations support the existence and stability of the R3m phase in our $ZrO_2$ films. The piezo response force microscopy (PFM) studies confirm the polar nature of the thin films, while the P-E hysteresis loops display saturation ferroelectric polarization up to about 20.2 $\mu C/cm^2$, and a coercive field of 1.5 MV/cm. The activation field estimated from polarization reversal characteristics is comparable to the coercive field obtained from polarization-electric field hysteresis loops.

## 2. Experimental Methods

$ZrO_2$ thin films with a thickness of 8 nm were grown by ion-beam sputter deposition (IBSD) technique onto 0.7 wt% Nb-doped (111)-$SrTiO_3$ substrates. The vacuum chamber was first evacuated down to a low pressure of $1\times10^{-6}$ mbar prior to the deposition. During the deposition, the substrate was kept at a temperature of 330 °C at a distance of 87.3 mm from the target. The gas pressure inside the chamber was maintained constant at $2.5\times10^{-4}$ mbar. A gas flow of 6.0 ml/min of Ar and 2.0 ml/min of $O_2$ was introduced into the ion beam gun and the atoms were ionized in the ion source with an RF-power of 120 W. The ion beam was further accelerated at 900 V and the ion beam current was maintained at 31 mA. After the deposition, the thin films were rapid thermal annealed in $N_2$ (6 mbar) at 700 °C, for 60 s.

The as-deposited thin films were structurally characterized by X-ray diffraction (XRD) with a Bruker D8 Discover diffractometer using Cu-Kα radiation (λ = 1.54056 Å).

Cross-section TEM specimens were prepared from the as-deposited samples by mechanical polishing down to ca. 30 μm, followed by ion milling in a Gatan PIPS machine at 4 kV accelerating voltage and 7° incidence angle. Low-voltage (2 kV) milling was used as final ion polishing stage in order to reduce the amorphous surface layer enveloping the specimen. Transmission electron microscopy observations were performed using a probe-corrected analytical high-resolution JEMARM 200F electron microscope operated at 200 kV.

To obtain the Annular Bright Field-scanning TEM (ABF–STEM) images we used the high-angle annular dark-field imaging detector with a camera length set at 20 cm corresponding to a collection angle range of 40–130 mrad. The condenser aperture of 30 µm was selected, corresponding to a beam convergence angle of 2.6 mrad.

DFT simulations of the polar rhombohedral polymorph of $ZrO_2$ were run by mimicking the approach employed in Ref. 8 to study its $Hf_{0.5}Zr_{0.5}O_2$ counterpart. All the calculations were carried out using the Perdew-Burke-Ernzerhof formulation for solids (PBE-sol) of the generalized gradient approximation to DFT [14] as implemented in the VASP code [15,16]. The atomic cores were treated within the projector augmented wave (PAW) approach [17], considering the following states explicitly: 4s, 4p, 4d and 5s electrons for Zr; 2s and 2p electrons for O. A plane wave basis set with a cut-off of 800 eV was employed to represent the electronic wave functions. In all our calculations, a periodically repeated 36-atom unit cell was considered, with approximate lattice parameters a = 7.197 Å, b = 7.197 Å, c = 8.822 Å, $\alpha = \beta = 90°$ and $\gamma = 120°$. Reciprocal space integrals were solved using a 4x4x3 Γ-centered Monkhorst-Pack K-point mesh [18] to sample the Brillouin zone. Atomic structures were fully relaxed until the residual forces fell below 0.01 eV/Å and residual stresses below 0.1 GPa. To simulate the effect of epitaxial strain, the in-plane lattice parameters were held fixed, while the out-of-plane lattice parameter and atomic positions were relaxed. The polarization was computed using the modern Berry phase theory of polarization [19].

Piezoresponse force microscopy was done using the scanning probe microscope NT-MDT Solver NTEGRA equipped with external lock-in amplifiers. We used commercial Tap190E-G probes with Cr/Pt coating, tip radius of 25 nm, resonance frequency ~151 kHz and spring constant k about 48 N/m. The topographic, piezo-response amplitude and phase images are edited via WSxM 5.0-9.0 software. All piezoresponse force microscopy and spectroscopy studies were done out-of-resonance at 21.1(1) kHz in order to decrease electrostatic responses with correspondent topographic crosstalk [20].

For the ferroelectric characterization, circular gold (Au) electrodes, having a diameter of 1 mm were deposited by thermal evaporation on the upper surface of the films. The ferroelectric hysteresis loops (P-E) were measured at room temperature with a modified

Sawyer-Tower circuit using a sinusoidal signal at 1 kHz. The polarization reversal characteristics have been studied by applying square pulses, and the corresponding current was measured across a resistance of 100 Ω connected in series with the sample.

## 3. Results and discussion

### 3.1. Microstructure analysis

Figure 1 shows a representative XRD pattern in the 20–42.5° 2θ-range, recorded at ambient conditions on the $ZrO_2$ thin film deposited onto (111)-oriented Nb:STO substrate. The XRD pattern exhibits the strongest Bragg peak approximately centered at $2\theta = 39.97°$, originating from the (111) planes of the $SrTiO_3$ substrate. One significant diffraction peak centered at $2\theta = 30.27 \pm 0.03°$ ($d = 2.95 \pm 0.01$ Å) is also observed. However, due to the broadening of the diffraction peak, it is impossible to unambiguously attribute this peak to either of the usual monoclinic, orthorhombic or tetragonal phases that are well-known for $ZrO_2$. The peak positions corresponding to the monoclinic, orthorhombic and tetragonal phases are marked in the figure for comparison [13,21,22]. Thus, to unravel the crystalline nature of the $ZrO_2$ films, TEM investigations were conducted.

Figure 2 represents a HRTEM image showing the $ZrO_2$ layer with the thickness of 8 nm. Here we can observe a mixing of crystalline (A1 and A3) and amorphous (A2) areas inside the $ZrO_2$ thin film. The fast Fourier transform (FFT) patterns (Fig. 2 right side) corresponding to the areas indicated in the HRTEM image show the nature of each of them (crystalline or amorphous) by the presence or absence of intensity peaks.

In order to determine the internal structure of the $ZrO_2$ thin film we tried to index the FFT patterns according to the most known phases: orthorhombic/tetragonal and monoclinic. However, the measured values of $d_{111}$ and $d_{11-1}$ are different, which implies that we have to search for a different structural phase. Hence, we considered the possibility that our $ZrO_2$ films might present a rhombohedral R3m phase akin to the one recently reported for $Hf_{0.5}Zr_{0.5}O_2$ films [8]. Thus, we combined TEM analysis with DFT calculations to reveal an indicative of the rhombohedral phase in the present $ZrO_2$ films rather than the monoclinic and orthorhombic phases commonly observed in constrained $ZrO_2$ films.

### 3.2 Density functional theory calculations

We ran first-principles simulations that mimicked those reported in Ref. 8, but considering $ZrO_2$ instead of HZO. Our results are summarized in Fig. 3. In our simulations, we mimicked epitaxial strain by imposing the in-plane lattice constants (see section 2) and letting all other structural degrees of freedom relax. We thus obtained a dependence of the $d_{111}$ inter-plane spacing as a function of the in-plane lattice constant, as shown in Fig. 3(a): naturally, larger $d_{111}$ values correspond to a stronger in-plane epitaxial compression. Additionally, we computed the polarization for the obtained configurations; the results are shown in Fig. 3(b) as a function of $d_{111}$.

These results confirm that, as HZO, $ZrO_2$ also presents a strongly polar polymorph whose polarization grows as both the in-plane compression and $d_{111}$ increase. Examples of the structures obtained from DFT, for representative $d_{111}$ values (2.94 Å and 3.21 Å), are provided as supplemental material (Table S1).

In order to determine the structural phase of the crystalline areas, diffraction patterns for the rhombohedral $ZrO_2$, with R3m space group, were simulated (Fig. 4b) using our DFT data, taking different $d_{111}$ values in the range represented in Fig. 3(a). The simulated diffraction patterns were compared with the FFT patterns (Fig. 4(a) and Fig. 5(c)) corresponding to the experimental ABF-STEM image in Fig. 5(a). The best fit between the experimental FFT pattern and the simulated diffraction pattern was obtained for the [-110] zone axis orientation and $d_{111}$=2.94 Å.

Further, the Annular Bright Field (ABF) imaging in STEM mode has been used for the direct visualization of the oxygen atoms around the zirconium atoms. The contrast in the ABF image was inverted for a better visibility of oxygen atoms. The ABF image of an area inside of the $ZrO_2$ thin film is presented in Fig. 5(a) and (b). The ABF image of $ZrO_2$ layer was compared with our DFT structural model (Fig. 5(d)) of the R3m phase (inset in Fig. 5(a)) with hexagonal lattice parameters a = b = 7.197 Å, c = 8.822 Å and α = β = 90°, γ = 120° (corresponding to $d_{111}$ = 2.94 Å in the trigonal description). Along the [-110] zone-axis of the $ZrO_2$ layer the oxygen columns are situated around the Zr columns as in the atomic model (Fig. 5(d)) and can be observed better in the zoomed area in Fig. 5(b). In addition, the O-Zr-O//O-Zr-O columns are collinear (green line in Fig. 5b), which rules out the possible existence of the R3 phase, where the loss of mirror symmetry would result in the loss of this collinearity [9]. This result represents the first experimental evidence of the R3m phase in $ZrO_2$ films. Interestingly, note that, in previous works on $Hf_{0.5}Zr_{0.5}O_2$ films, it was not possible to distinguish between the R3m or R3 symmetries [8,11].

Usually, it is observed that the rhombohedral R3m phase is stabilized by a combination of the large surface energy induced internal pressure of the nanoparticles and the substrate-imposed compressive strain. In the present case also, the growing crystallites are subjected to a large epitaxial compressive strain that elongates the cubic unit cell along the out-of-plane [111] direction, inducing rhombohedral symmetry with a polar unit cell [8].

Therefore, the Nb:STO (111)-oriented substrate provides a suitable template for the growth of the rhombohedral R3m phase of $ZrO_2$, without the formation of an in-plane tensile-strained tetragonal phase, as observed when $Hf_{0.5}Zr_{0.5}O_2$ films are deposited on (001)-oriented $La_{0.7}Sr_{0.3}MnO_3/SrTiO_3$ substrates [8]. (Yet, we do obtain amorphous $ZrO_2$ regions intercalated with the crystalline rhombohedral ones.) The IBSD technique was combined with RTA that allows the growth of rhombohedral R3m phase $ZrO_2$ at lower temperatures, when compared to the conditions used to grow rhombohedral R3m phase $Hf_{0.5}Zr_{0.5}O_2$ films [8]. The lower temperatures used in this work are also very advantageous from the technological point-of-view and also responsible for the existence of amorphous $ZrO_2$ regions between the crystalline ones.

### 3.3 Ferroelectricity in rhombohedral $ZrO_2$ films

The topographic scan shown in Fig. 6(a), obtained in a representative 2x2 $\mu m^2$ region of the film, reveals a very homogeneous and smooth surface with an average roughness of only ~0.12 nm, which does not allow to distinguish grain boundary contours. Detailed amplitude and phase piezo-response scans of the same area shown in Figs. 6(b)-(f) reveal clear contrast regions with no apparent topological correlation. The profile of the domains with an in-plane component of polarization configure some typical ~100 by 300 nm isolated regions; the profile of domains with the out-of-plane component of polarization reveals more diffuse ~200 by 300 nm isolated regions, also exhibiting a relatively weaker phase contrast in comparison to the in-plane phase signal.

In Fig. 6, the PFM amplitude and phase scans are edited by considering two levels in the scale between the maximum and minimum values. These contours enable to enhance the contrast between three different regions conforming the domains having opposite oriented polarization components (red/blue) and without contrast (green). In $ZrO_2$ thin films, contrary to conventional ferroelectrics and relaxors [24], it is observed that domains with opposite orientations (red/blue) are not in direct contact with each other, and are in fact entangled and surrounded by dendritic-like non-polarized regions (green). These non-polarized regions can be associated with amorphous $ZrO_2$, while the polar ones can be associated with the rhombohedral R3m phase, as shown by TEM analysis.

Local out-of-plane amplitude and phase piezoresponse loops under +-15 Vdc ramps depicted in Fig. 6(d) have well-defined, reversible and reproducible cycles, corroborating the ferroelectric nature of this $ZrO_2$ film phase. The cycles have an -1 V offset with coercive fields set at -4.5 and +2.5 V, while saturation fields set at -7 and +5 V. This asymmetry between negative and positive bias fields is due to the asymmetry between the film top and bottom electrodes, in this case between the Cr/Pt tip and the Nb/STO substrate. By comparison to the response of a $LiNbO_3$ reference sample "PFM3" [24] measured by PFM under the same tip and conditions [25], the $d_{33}$ response of the $ZrO_2(111)$ film surface is roughly estimated of the order of 10 pm/V, corresponding to a local out-of-plane polarization with a component of at least >1 $\mu C/cm^2$ in the 8 nm thin film.

To further investigate the nature of the non-responsive "green" regions observed in Fig. 6, the two amplitude scans were combined, edited and analyzed to reveal in Fig. 7 a composed image, where the distribution of in-plane and out-of-plane domains is now represented in terms of their polarization components modulus as blue and red regions respectively. Moreover, the image clearly exposes the percolative spatial distribution and confirms a non-polarizable response of the "white" zones common to both scans, which can be associated to the observed amorphous $ZrO_2$ regions in the film (Fig. 2).

Figure 8 displays the room temperature electric field dependent polarization-electric field (P–E) hysteresis loops of the Nb:STO/$ZrO_2$/Au film capacitors. It is possible to observe from the figure that, as the voltage increases, the P–E loop saturates. As the electric field increases from 2.1 MV/cm to 5.3 MV/cm, both the saturation polarization ($P_s$) and the remnant polarization ($P_r$) increase. The extracted $P_s$, $P_r$ and coercive field ($E_c$) values averaged from 50 measurements are shown in Table 1 for the different applied electric fields. For an electric field of 5.3 MV/cm, a well-saturated hysteresis loop was obtained with a $P_s$ of 20.2 $\mu C/cm^2$, $P_r$ of 10.8 $\mu C/cm^2$ and $E_c$ of 1.5 MV/cm. The obtained values for $P_r$ and $P_s$ are comparable to those reported in the literature for orthorhombic $ZrO_2$ film capacitors, in which a wake-up cycle was needed [13,26,27]. In addition, the polarization values are lower than the ones found in rhombohedral 5-nm $Hf_{0.5}Zr_{0.5}O_2$ films [8], while the coercive field is significantly lower (more than 3 times less), which is beneficial for technological applications. Reduction of the switching voltages of these materials is considered a key step for memory applications [28].

Note that the measured $P_s$ would correspond to a $d_{111}$ inter-layer spacing of about 3.2 Å according to our DFT simulations (see Fig. 3(a)). In contrast, when we consider the diffraction and HRTEM data, the best match between the experiment and theory corresponds to the structure obtained for $d_{111}$ = 2.94 Å, which has a much smaller (comparatively negligible, see Fig. 3(a)) polarization value associated with it. It is worth noting that a similar discrepancy between the theory and experimental data affected the results for $Hf_{0.5}Zr_{0.5}O_2$ in Ref. 8. Hence, it is clear that the idealized $ZrO_2$ films considered in our simulations (where the peculiarities of a thin film are introduced by just a simple elastic constraint) do not capture the complexity of the real material, and that other factors must be responsible for the experimental observation of the large remnant polarizations for moderate $d_{111}$ interplanar distances. One of these factors could be the thickness of the film. As shown in Ref. 8 for $Hf_{0.5}Zr_{0.5}O_2$ films, a similarly large $d_{111}$ spacing, comparable to the one suggested by the DFT results, was observed for a 1.5 nm thick film. With further increase in thickness, the observed $d_{111}$ value tends to decrease while the measured polarization is still relatively large. Hence, a more detailed comparison between theory and experiment should take into account these and other factors (e.g., the polycrystallinity of our films), and remains for future work. Nevertheless, at present, the ability to match our structural data, together with mounting related evidence for related-compound $Hf_{0.5}Zr_{0.5}O_2$, make us confident that our simulated rhombohedral $ZrO_2$ is a fair representation of the experimentally-obtained structure.

The time dependence polarization reversal characteristics of the Nb:STO/$ZrO_2$/Au film capacitors were also investigated. The bell shape curve in polarization current arises due to domains reversal and this behavior is similar to the one already observed in HZO orthorhombic films [29]. The maximum value in polarization current ($i_m$) occurs at a time ($t_m$), which corresponds to the situation where most of the domains are reversed. The effect of pulse amplitude on the time dependence of polarization current transients of Nb:STO/$ZrO_2$/Au film capacitors is shown in Fig. 9(a). As the applied electric field increases from 2.1 to 5.3 MV/cm, the $i_m$ value increases from 5.35 to 15.56 mA, whereas the $t_m$ values decrease from 20.2 to 15.0 ns. The $i_m$ values exhibit an exponential decrease with E according to [30]

$$i_m = i_0 \exp(-\alpha_i/E), \qquad (1)$$

where $\alpha_i$ is the activation field. The semi-log plot of $i_m$ versus $1/E$ is shown in the inset of Fig. 9(a). The activation field was estimated to be 3.9 MV/cm. This value is comparable to the coercive field obtained from the P-E loops (1.5 MV/cm). As discussed by Lee et al., in conventional ferroelectrics the coercive field is about one tenth of the activation field, while in binary oxides, such as $HfO_2$, the coercive field is unusually large and even comparable with the activation field [31].

The polarization during the switching ($P_s$) can be estimated from the area (equal to charge Q) under the transient curves by using the relation [30]:

$$Q = 2P_S A \qquad (2)$$

where A is the area of the electrode. The saturation polarization obtained for each applied electric field is shown in Fig. 9(b) and is found to be in a good agreement with the values obtained from the P-E loops also plotted in the same Figure.

Therefore, through the PFM, P-E loops and time dependent polarization reversal we could prove the ferroelectric nature of rhombohedral R3m epitaxially-strained $ZrO_2$ thin films. Remarkably, and in contrast to previous investigations, the ferroelectric behavior was observed without any wake-up cycling.

## 4. Conclusions

Rhombohedral R3m epitaxially-strained $ZrO_2$ thin films were deposited on (111)-oriented $Nb:SrTiO_3$ substrates by ion-beam sputtering deposition technique. The experimental structural characterization, combined with DFT calculations, revealed, for the first time, the presence of a polar rhombohedral R3m phase in $ZrO_2$ thin films. Ferroelectric properties were investigated at the nano- and macro level and revealed a large saturation polarization of 20.2 $\mu C/cm^2$, with a coercive field of 1.5 MV/cm. Most importantly, the ferroelectric behavior is observed in the as-grown sample, without any need of performing any wake-up cycling. In addition, the coercive field was found to be comparable with the estimated activation field obtained from the time dependent polarization switching.

Therefore, we believe the present work might inspire new investigations in nanosized binary oxide thin films that can show a ferroelectric behavior and can be used in the next-generation memory devices.

**Acknowledgements**

This work was supported by: (i) the Portuguese Foundation for Science and Technology (FCT) in the framework of the Strategic Funding Contract UIDB/04650/2020 (ii) DST-SERB, Govt. of India through Grant Nr. ECR/2017/00006 (iii) Project NECL - NORTE-01-0145-FEDER-022096 and Project UID/NAN/50024/2019. This work was also developed within the scope of the project CICECO-Aveiro Institute of Materials, refs. UIDB/50011/2020 and UIDP/50011/2020, financed by national funds through the Portuguese Foundation for Science and Technology/MCTES.

R. F. N., M. C. I. and C. G. acknowledge the financial support from the Romanian Ministry of Education and Research within the project PN-III-P4-ID-PCCF2016-0047, contract no. 16/2018. Work at LIST was supported by the Luxembourg National Research Fund through projects PRIDE/15/10935404 "MASSENA" (S.D.) and INTER/ANR/16/11562984 "EXPAND" (H.A. and J.Í.). The authors acknowledge the CERIC-ERIC Consortium for access to experimental facilities and financial support under proposal 20192055. The authors would also like to thank José Santos for technical support in the Thin Film Laboratory at CF-UM-UP.

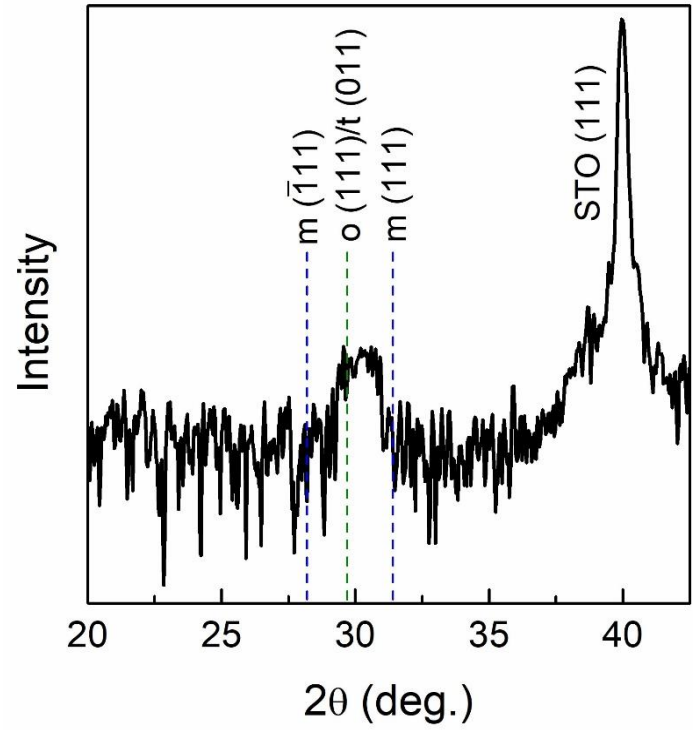

Figure 1. X-ray diffraction patterns of the ZrO$_2$ film deposited on Nb:STO (111) substrate. The intensity is in the log scale.

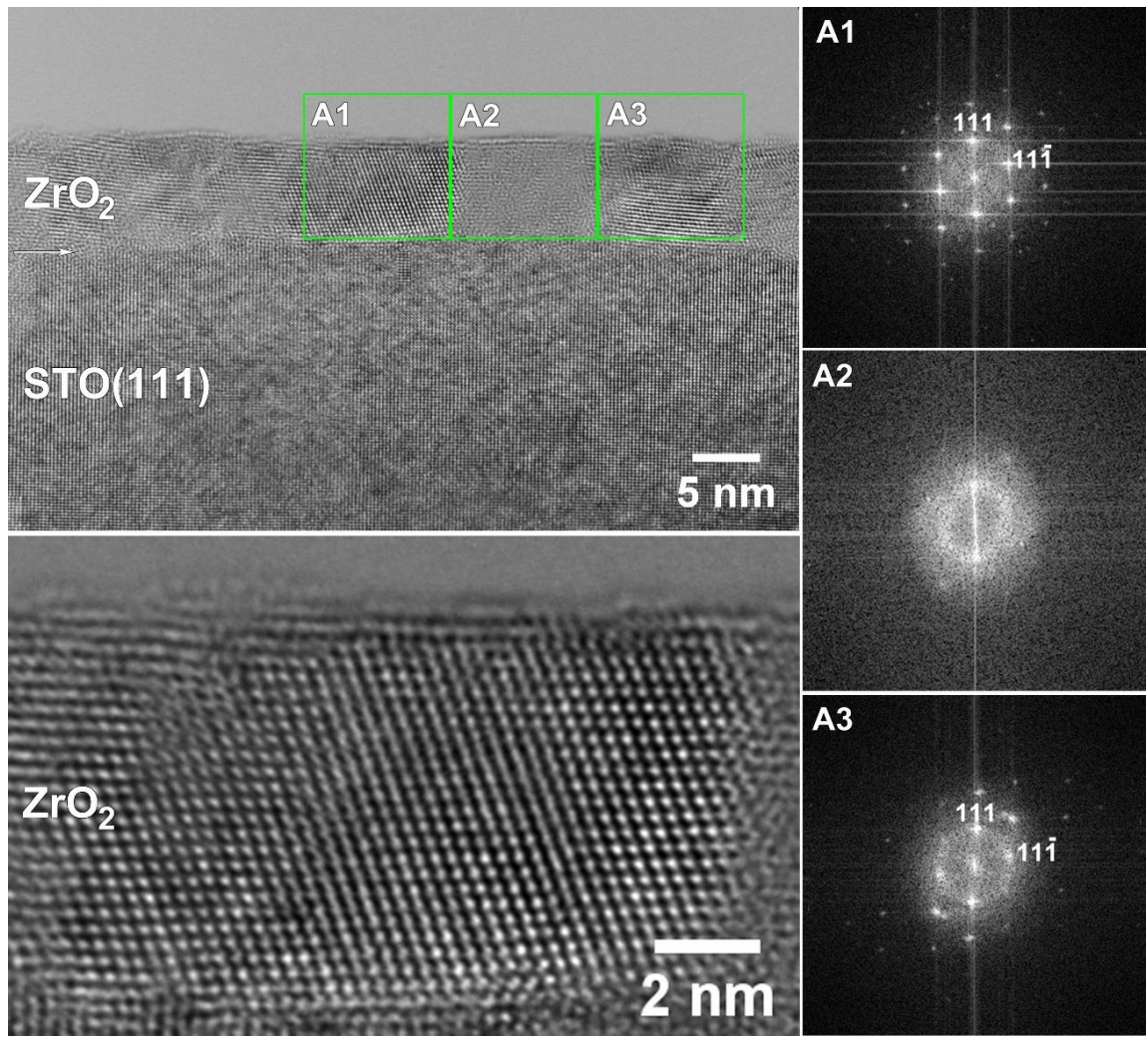

Figure 2. HRTEM images of the ZrO$_2$ layer deposited on STO (111) substrate showing crystalline areas with rhombohedral phase (e.g. A1 and A3) mixed with amorphous areas (e.g. A2). The arrow in the first figure marks the interface between the STO and ZrO$_2$. In the right side the FFT patterns corresponding to the areas indicated in HRTEM image show the nature of each area (crystalline/amorphous).

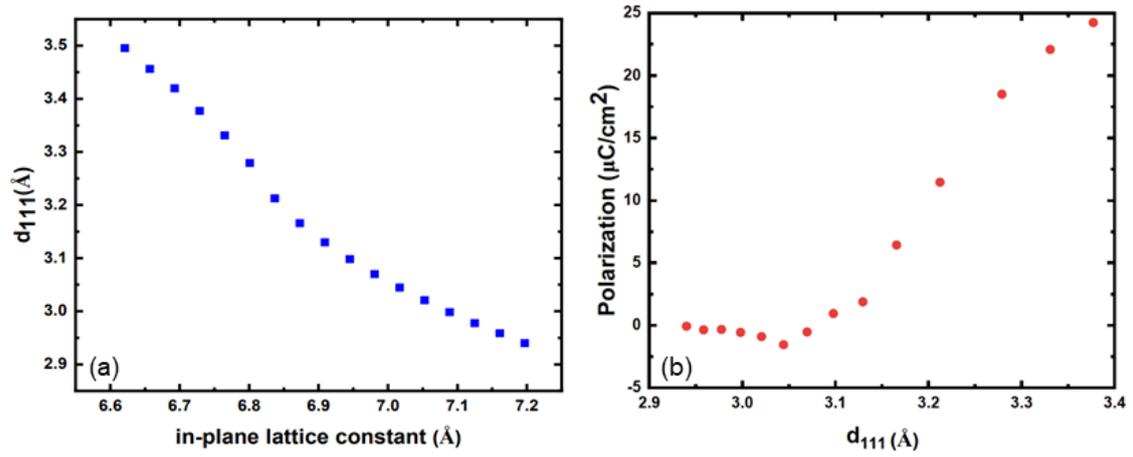

Figure 3. (a) Computed evolution of the $d_{111}$ inter-plane spacing of the R3m structure of $ZrO_2$ as a function of the imposed in-plane lattice constant. (b) Computed out-of-plane polarization as a function of $d_{111}$.

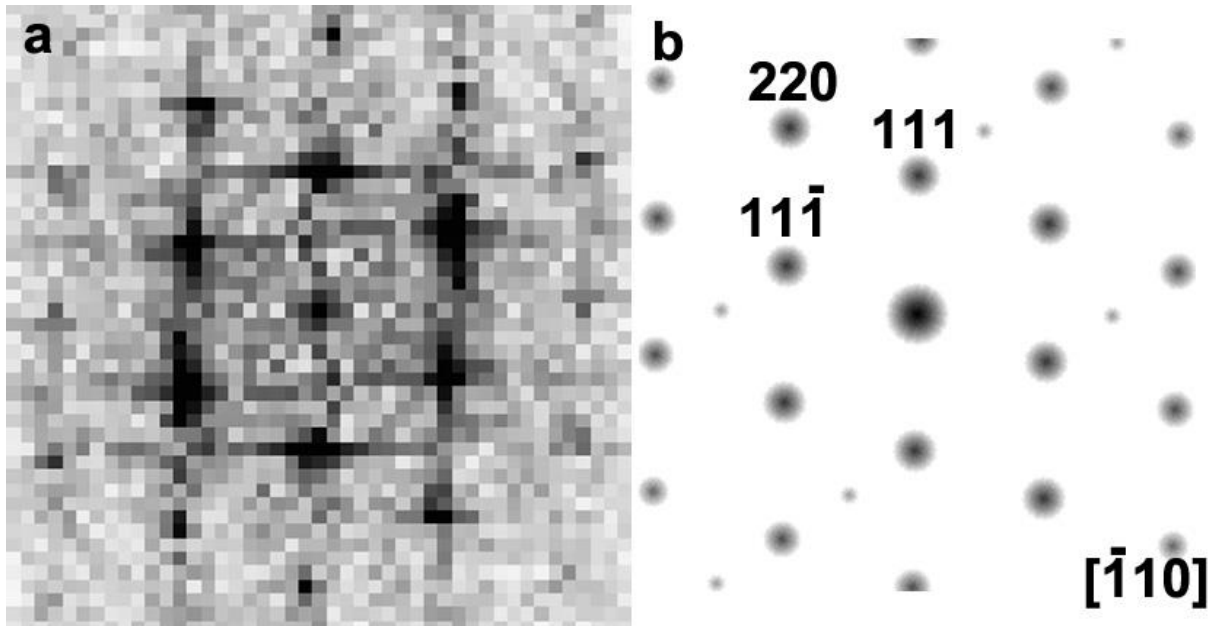

Figure 4. (a) FFT pattern corresponding to the HRTEM image from Figure 2(a) and (b) simulated SAED pattern of rhombohedral ZrO$_2$ (space group R3m) in trigonal system and [-110] zone-axis of ZrO$_2$ using the lattice parameters predicted by the DFT model.

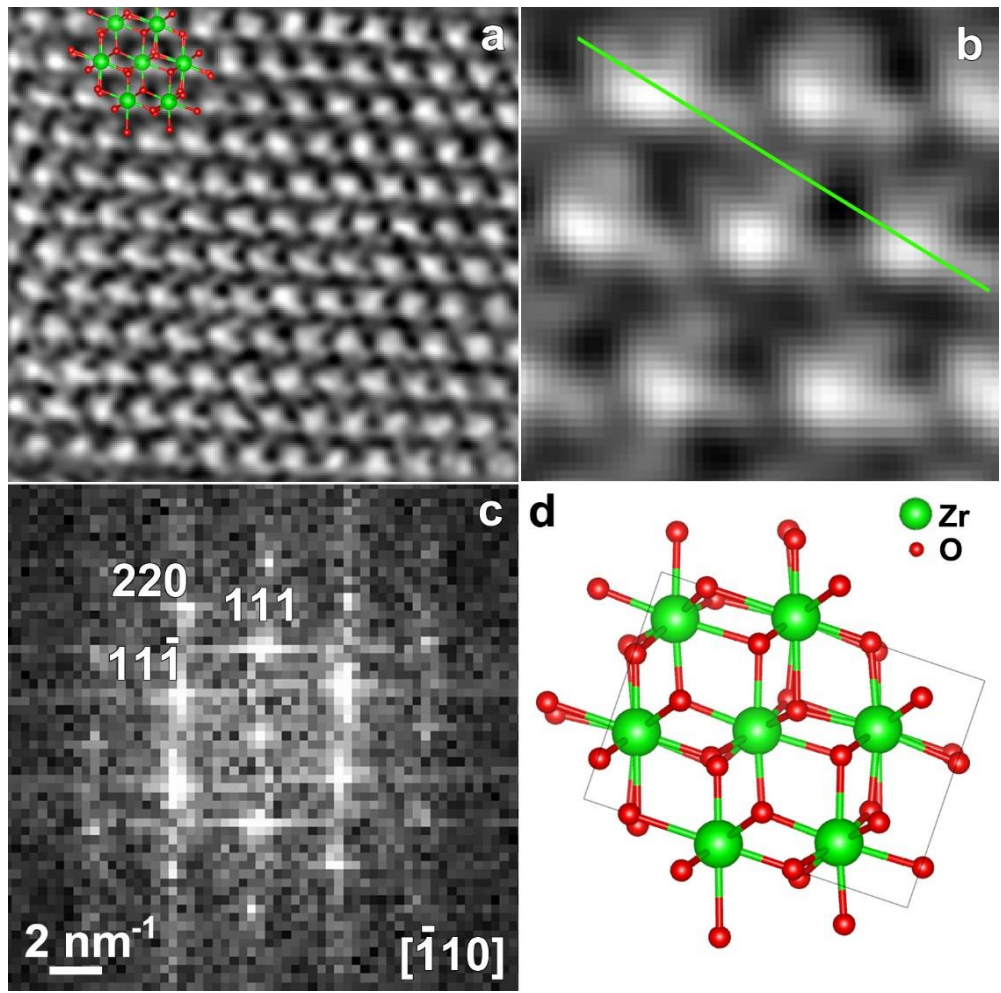

Figure 5 (a) Annular Bright Field - STEM image of $ZrO_2$ structure (inset – atomic model of $ZrO_2$); (b) a detail extracted from ABF-STEM image (a) showing the atomic arrangement of zirconium and oxygen. The green line is a guide to the eye showing that the O-Zr-O//O-Zr-O columns are collinear; (c) FFT pattern corresponding to the ABF-STEM image (a) in [-110] orientation; (d) atomic model of $ZrO_2$ with R3m space group in trigonal system and [-110] orientation.

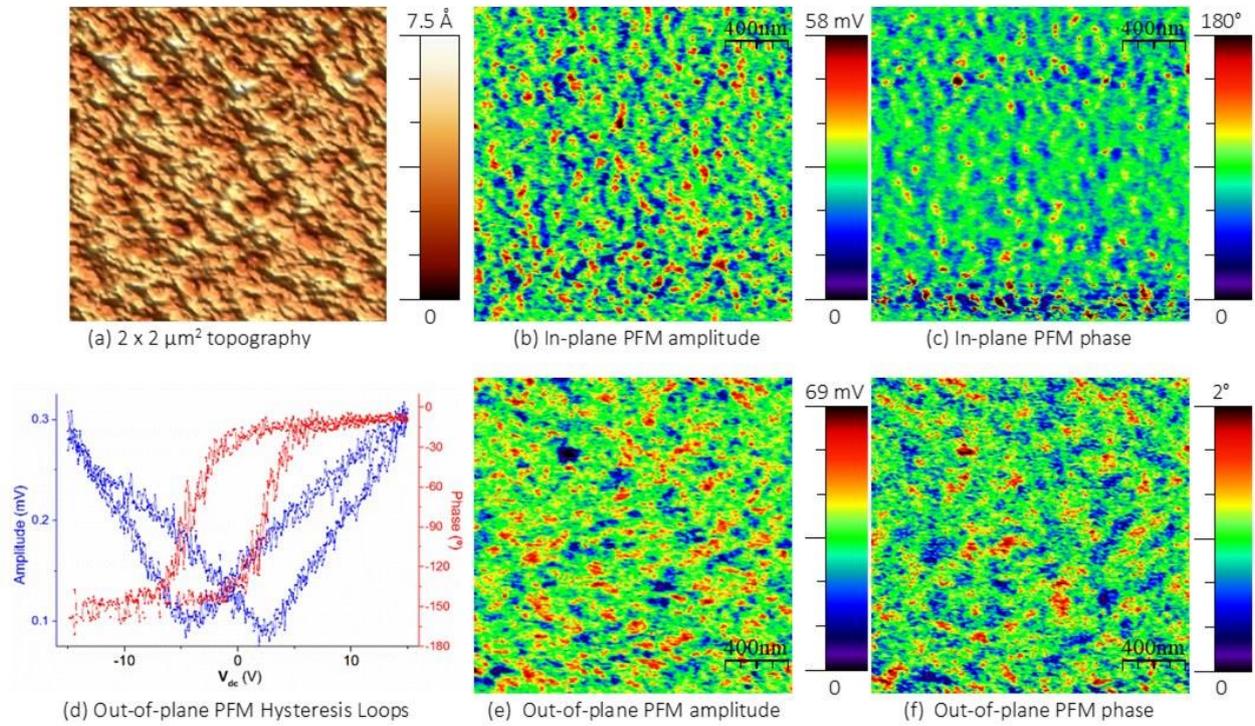

Figure 6 ZrO$_2$ thin film 2x2 μm$^2$ surface (a) representative topographic scan and respective piezo-response scans of in-plane (b) amplitude and (c) phase and out-of-plane (d) local hysteresis loops (e) amplitude and (f) phase.

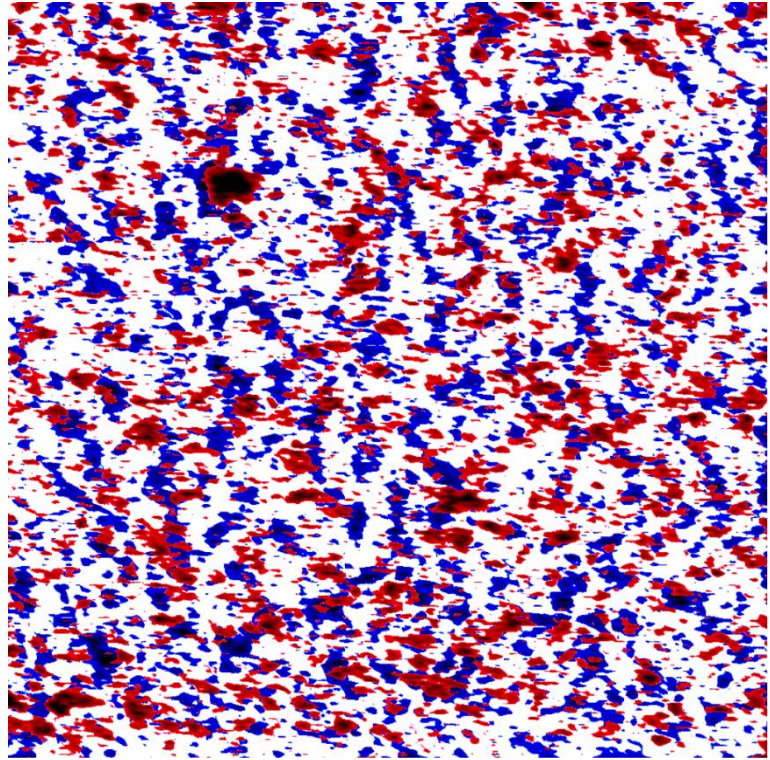

Figure 7 An image combining in-plane (blue) with out-of-plane (red) PFM amplitude modulus from ZrO$_2$ thin film (2x2 µm$^2$) surface enabling to distinguish common non-responsive regions (white).

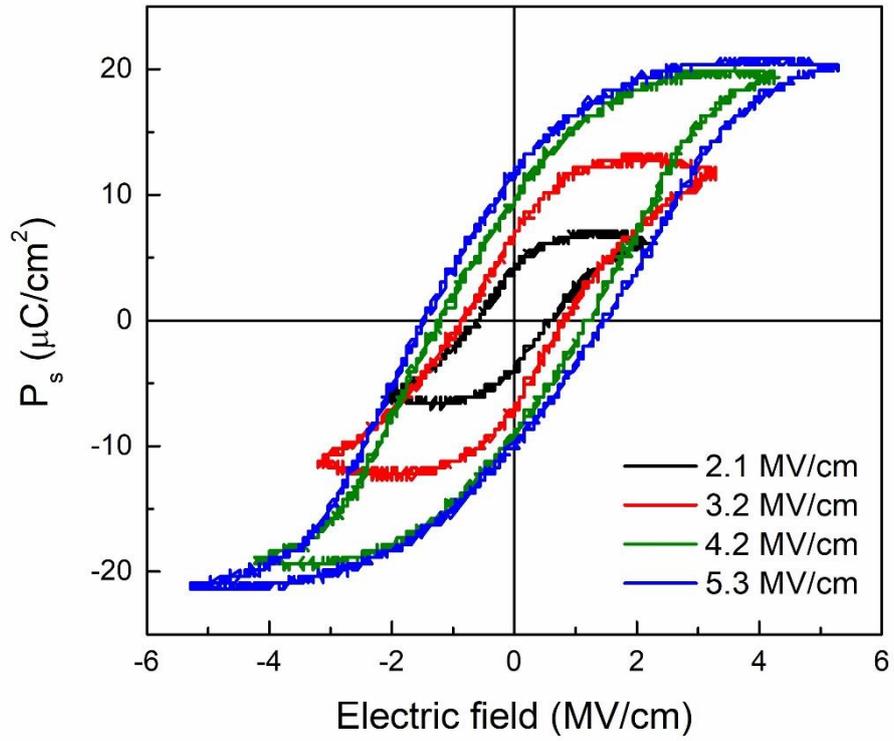

Figure 8. Electric field dependent P–E hysteresis loops of the Nb:STO/ZrO$_2$/Au film capacitors.

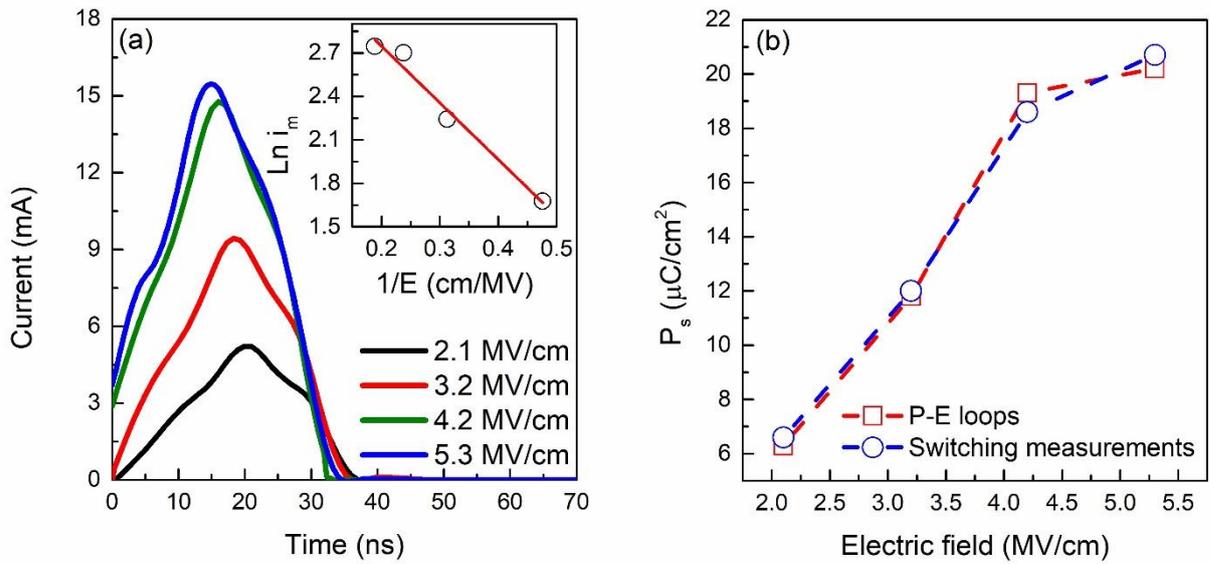

Figure 9 (a) Time dependence polarization reversal characteristics of the Nb:STO/ZrO$_2$/Au film capacitors. The inset shows the semi-log plot of $i_m$ versus 1/E. (b) Electric field dependence of the saturation polarization values obtained from the P–E hysteresis loops and switching measurements.

Table 1. Average $P_s$, $P_r$ and $E_c$ values as a function of applied electric field obtained for the Nb:STO/ZrO$_2$/Au film capacitors.

| Electric field (MV/cm) | $P_s$ (µC/cm$^2$) | $P_r$ (µC/cm$^2$) | $E_c$ (MV/cm) |
|---|---|---|---|
| 2.1 | 6.3  | 4.2  | 0.6 |
| 3.2 | 11.8 | 6.9  | 0.8 |
| 4.2 | 19.3 | 9.2  | 1.2 |
| 5.3 | 20.2 | 10.8 | 1.5 |

# Supplementary information

# Ferroelectricity in epitaxially strained rhombohedral ZrO$_2$ thin films

J. P. B. Silva[a], R. F. Negrea[b], M. C. Istrate[b], S. Dutta[c], H. Aramberri[c], J. Íñiguez[c,d], F. G. Figueiras[e], C. Ghica[b], K. C. Sekhar[f], A. L. Kholkin[g]

[a]Centre of Physics of University of Minho and Porto (CF-UM-UP), Campus de Gualtar, 4710-057 Braga, Portugal
[b]National Institute of Materials Physics, 105 bis Atomistilor, 077125 Magurele, Romania
[c]Materials Research and Technology Department, Luxembourg Institute of Science and Technology (LIST), 5 avenue des Hauts-Fourneaux, L-4362, Esch/Alzette, Luxemburg
[d]Department of Physics and Materials Science, University of Luxembourg, Rue du Brill 41, L-4422 Belvaux, Luxembourg
[e]IFIMUP & Department of Physics and Astronomy, Sciences Faculty, University of Porto, Rua do Campo Alegre, 687, 4169-007 Porto, Portugal
[f]Department of Physics, School of Basic and Applied Science, Central University of Tamil Nadu, Thiruvarur-610 101, India
[g]Department of Physics, CICECO-Aveiro Institute of Materials, University of Aveiro, 3810-193 Aveiro, Portugal

| ZrO$_2$ **R3m (No. 160)** for d$_{111}$=2.94 Å |||||
| a=b=7.20 Å; c=8.82 Å; α=90˚; β=90˚; γ=120˚ |||||
| Zr | 9b | 0.833290 | 0.166710 | 0.250990 |
| Zr | 3a | 0.000000 | 0.000000 | 0.583940 |
| O | 9b | 0.149880 | 0.850120 | 0.159130 |
| O | 9b | 0.487070 | 0.512930 | 0.327920 |
| O | 3a | 0.000000 | 0.000000 | 0.859740 |
| O | 3a | 0.000000 | 0.000000 | 0.353410 |
| **ZrO$_2$ R3m (No. 160)** for d$_{111}$=3.21 Å |||||
| a=b=6.84 Å; c=9.64 Å; α=90˚; β=90˚; γ=120˚ |||||
| Zr | 9b | 0.830540 | 0.169460 | 0.261890 |
| Zr | 3a | 0.000000 | 0.000000 | 0.561520 |
| O | 9b | 0.151020 | 0.848980 | 0.155060 |
| O | 9b | 0.483950 | 0.516050 | 0.329460 |
| O | 3a | 0.000000 | 0.000000 | 0.861500 |
| O | 3a | 0.000000 | 0.000000 | 0.348960 |

Table S1. Atomic structures of rhombohedral ZrO$_2$, obtained from DFT, for two different values of the d$_{111}$ inter-planar distance, i.e., 2.94 Å and 3.21 Å. These structures correspond to imposing in-plane lattice constants a=b of 7.20 Å and 6.84 Å, respectively.